\documentclass[12pt,preprint]{aastex}

\shorttitle{The CHARA Array}
\shortauthors{ten Brummelaar et al.}

\begin{document}

\title{First Results from the CHARA Array. II. A Description of the Instrument}


\author{T.A. ten Brummelaar}
\affil{The CHARA Array, Mt. Wilson Observatory, Mt. Wilson, CA 91023}
\email{theo@chara-array.org}

\author{H.A. McAlister}
\affil{Center for High Angular Resolution Astronomy, Georgia State University, P.O. Box 3969, Atlanta, GA 30302-3969}
\email{hal@chara.gsu.edu}

\author{S.T. Ridgway}
\affil{National Optical Astronomy Observatories, P.O. Box 26732, Tucson, 
AZ 85726}
\email{ridgway@noao.edu}

\author{W.G. Bagnuolo Jr.}
\affil{Center for High Angular Resolution Astronomy, Georgia State University, P.O. Box 3969, Atlanta, GA 30302-3969}
\email{bagnuolo@chara.gsu.edu}

\author{N. H. Turner, L. Sturmann, J. Sturmann, D.H. Berger, C.E. Ogden, 
R.~Cadman}
\affil{The CHARA Array, Mt. Wilson Observatory, Mt. Wilson, CA 91023}
\email{nils@chara-array.org, sturmann@chara-array.org, judit@chara-array.org,
berger@chara-array.org, ogden@chara-array.org, cadman@chara-array.org}

\author{C.H. Hopper}
\affil{Department of Physics and Astronomy, Georgia State University, 
P.O. Box 4106, Atlanta, GA 30302-4106}
\email{chopper@gsu.edu}

\and 

\author{M.A. Shure\altaffilmark{1}}
\affil{Center for High Angular Resolution Astronomy, Georgia State University, P.O. Box 3969, Atlanta, GA 30302-3969}
\email{m\_a\_shure@yahoo.com}

\altaffiltext{1}{Currently at the Space Systems Division of ITT industries, Inc.
P.O. Box 3700, Fort Wayne, IN 46801.}

\vskip 0.5in

\begin{abstract}
The CHARA Array is a six 1-m telescope optical/IR interferometric array
located on Mount Wilson California, designed and built by the Center for
High Angular Resolution Astronomy of Georgia State University. In this
paper we describe the main elements of the Array hardware and software 
control systems as well as the data reduction methods currently being
used. Our plans for upgrades in the near future are also described.
\end{abstract}

\keywords{atmospheric effects --- instrumentation: high angular resolution --- 
instrumentation: interferometers --- techniques: high angular resolution --- 
techniques: interferometric --- telescopes}

\section{Introduction}

Georgia State University's Center for High Angular Resolution Astronomy 
(CHARA) has designed and built an optical/near-IR interferometric array on 
the grounds of Mount Wilson Observatory. This is the second of two papers 
concerning the CHARA Array. The first scientific results are presented in 
Paper I \citet{Regulus} of this series, while in this paper we will provide 
the technical information and performance background of the instrument. 

The CHARA Array consists of six 
1-m aperture telescopes arranged in a Y-shaped configuration 
yielding 15 baselines ranging from 34 to 331-m as well as 10 possible
phase closure measurements. This permits 
limiting resolutions for stellar diameter 
measurements, conservatively defined in terms 
of reaching the first null in visibility, of 1.6 and 0.4-mas 
(milli-arcseconds) in the K and V bands, respectively.

The major elements of the Array consist of light collecting telescopes, 
vacuum light beam transport tubes, optical path length delay lines, 
beam management systems, and beam combination systems. Superimposed on all 
these elements is a beam alignment system and, of course, an overall 
control system.  In this paper we will describe the main subsystems of
the Array, including the control system and data reduction methodology.
More in-depth technical information can be found in some of
our previous publications \citep{SPIE00, SPIE02, SPIE04} as well as in 
numerous internal technical reports available at CHARA's
website\footnote{http://www.chara.gsu.edu/CHARA/techreport.html}. 

\section{Site Considerations}

The Array is located on the grounds of the Mount Wilson Observatory just
north of Los Angeles, California. This well known observatory,
administered by the Mount Wilson Institute\footnote{http://www.mtwilson.edu} 
under an agreement
with the Carnegie Institution of Washington, is also the site
of several other high angular resolution experiments such as the
Berkeley Infrared Spatial Interferometer \citep{ISI}, working at 10-$\mu$m,
and the University of Illinois Seeing Improvement System \citep{UnISIS}, 
a laser guide star program. Active research also continues on the 
100-inch Hooker telescope and at the 60-ft and 150-ft solar towers.

Alternative locations in Arizona and New Mexico were considered 
for the CHARA Array, and Mount Wilson was ultimately chosen on the basis 
of its reputation
for excellent astronomical seeing \citep{Buscher_A,Buscher_B}, the large number of clear nights,
as well as on advantageous
cost and logistical aspects which largely centered around the
existing infrastructure. 
Light pollution from the city of Los Angeles has negligible effect on the 
Array's potential due to the extremely small field of view of the instrument. 
We have seen no sign of vibrations on the site other than those caused by our
own air conditioning compressors and the occasional
construction work on or near the mountain \citep{TR42}.
 
\section{Array Configuration on Mt. Wilson}

\begin{figure}
\centerline{\includegraphics[scale=0.70]{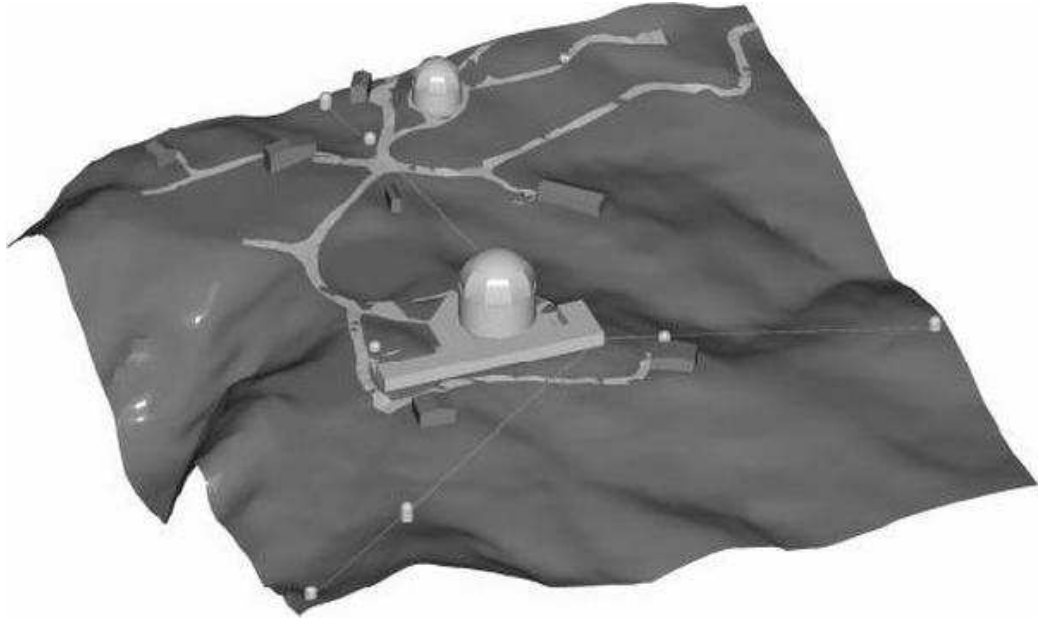}}
\vskip 0.1in
\centerline{\includegraphics[scale=0.60]{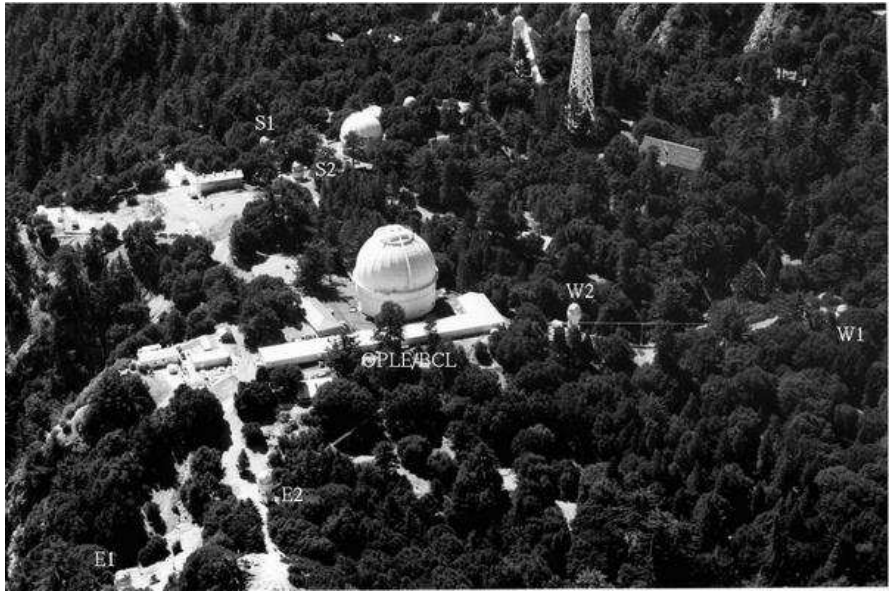}}
\caption{Top: Output of the computer model of the mountain showing the 
layout of the CHARA Array. This is a view from the north looking south.
The 100-in telescope dome is in the center with the L-shaped delay line 
and beam combining facility
building directly behind it.
Bottom: A photograph of the site taken from the north looking south. 
The 100-in dome is clearly visible behind the OPLE/BCL building.
\label{fig_overview}}
\end{figure}

The CHARA Array is built in a non-redundant 
Y configuration with two telescopes located along
each of the three arms of the interferometer. A great deal of attention
was paid to having as little impact on the existing structures and
trees along the Array arms
as possible, and while some road work was required, no existing structures and
only a handful of trees had to be removed to fit the Array on the
mountain. Figure \ref{fig_overview} shows two overviews of the mountain.
One is the output of the computer model of the mountain developed early
in the project \citep{TR48} to help us position the Array. This model
includes all existing roads, buildings and trees and allowed us to
experiment with various locations for the telescopes, buildings and light pipes 
and study their impact on the mountain. The second part of Figure
\ref{fig_overview} shows a photograph of the actual construction as it
was in 2000. All six domes and the central beam synthesis facility 
are visible in this picture as well as the vacuum light tubes from the east and west arms of the Array.

\begin{figure}
\centerline{\hbox{
\includegraphics[angle=90,scale=0.3]{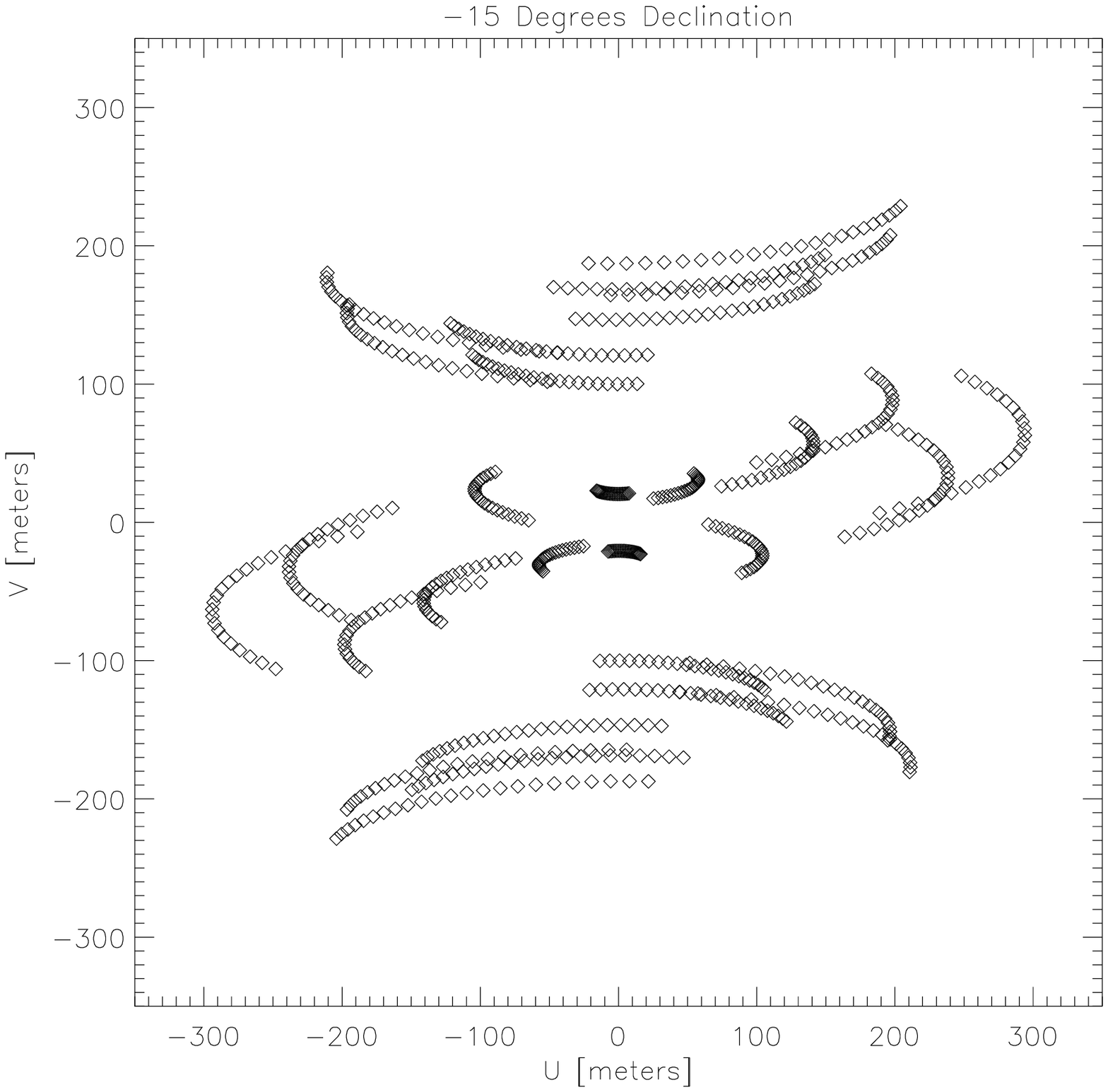} 
\hskip -0.5in
\includegraphics[angle=90,scale=0.3]{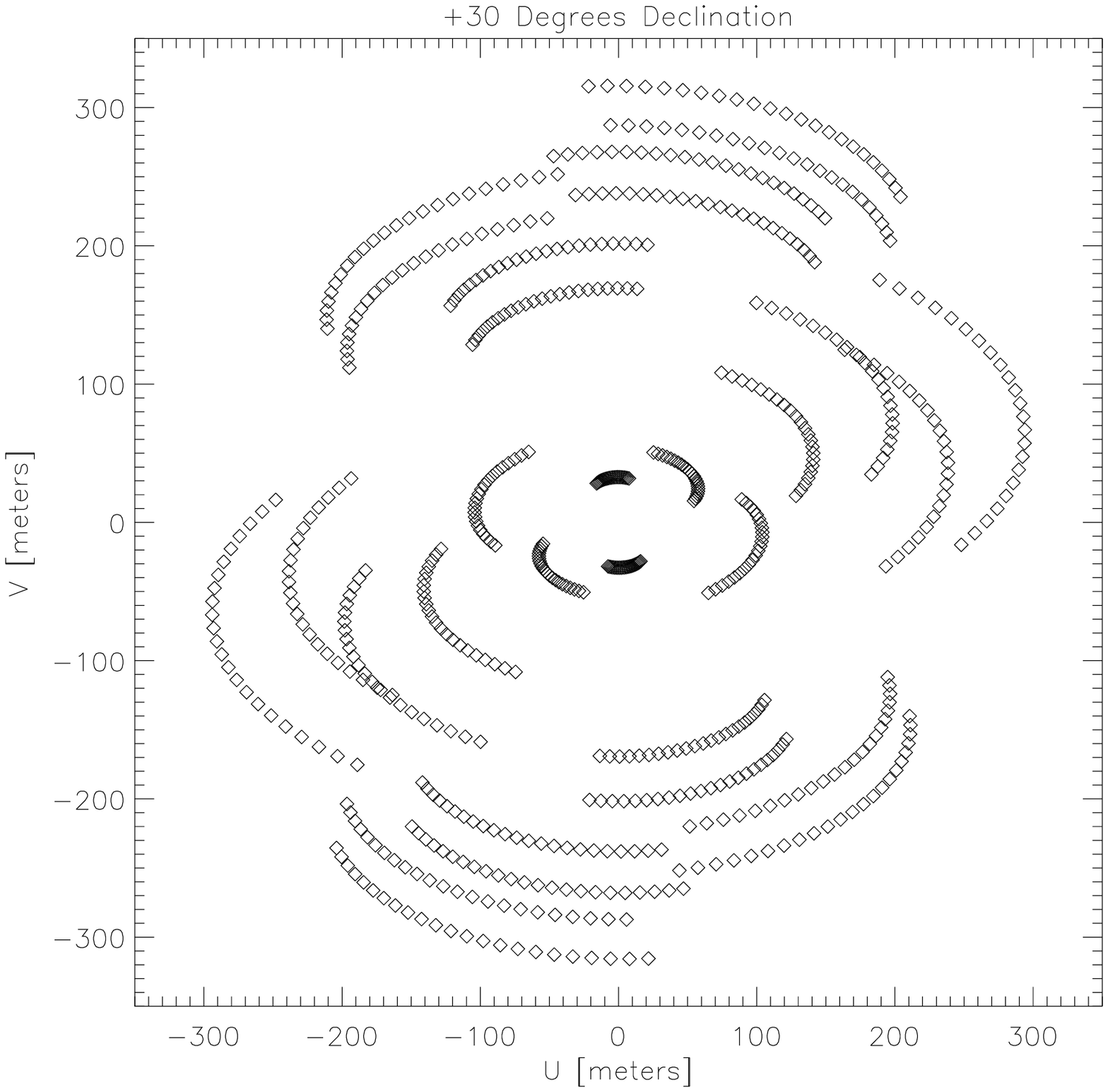}}}
\vskip 0.1in
\centerline{\includegraphics[angle=90,scale=0.3]{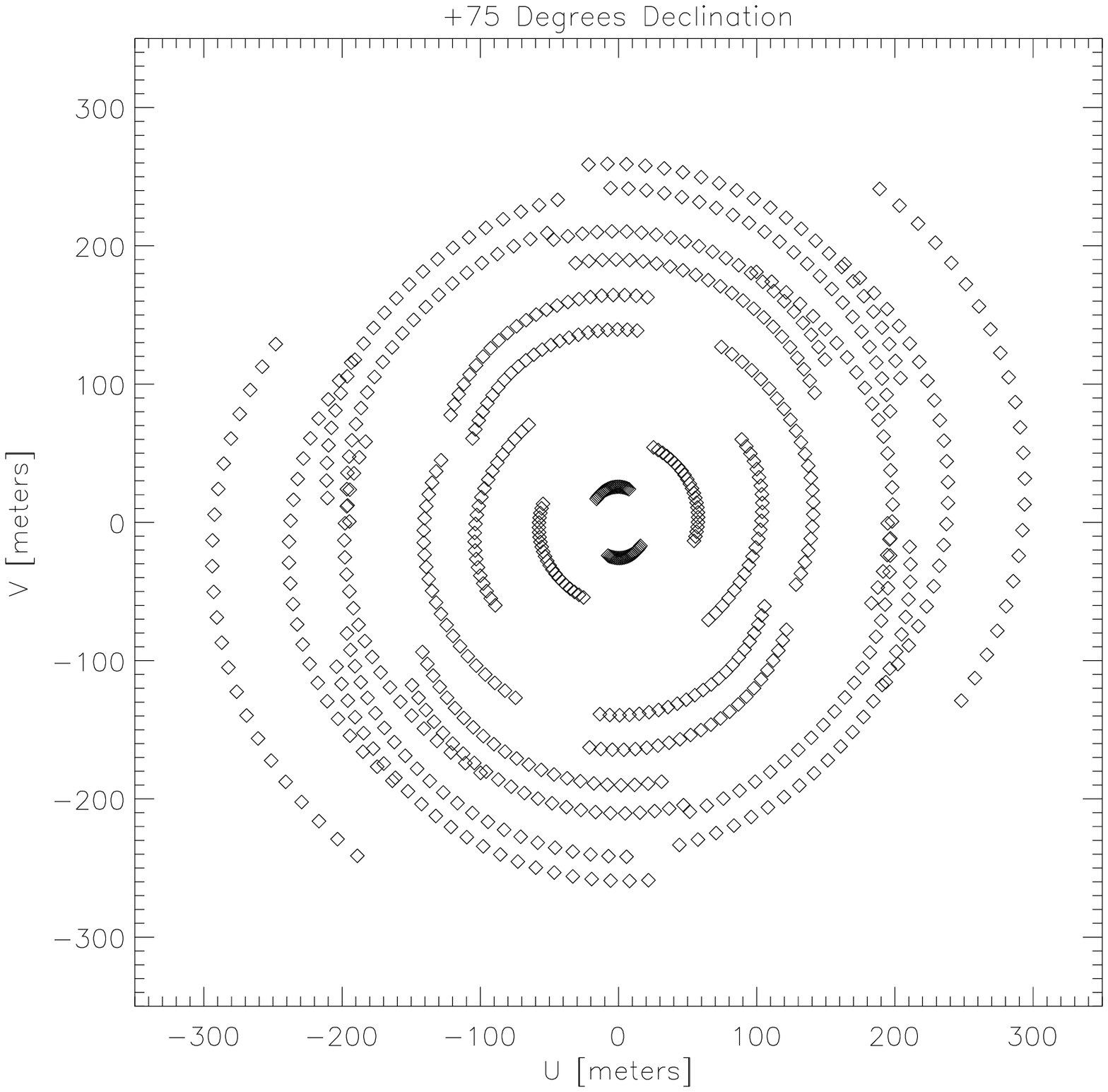}}
\caption{UV coverage of the CHARA Array for declinations -15$^{\circ}$ (top left), +30$^{\circ}$  (top right) 
and +75$^{\circ}$  (bottom). Each plot shows the coverage 
resulting from all six telescopes within three hours either side of transit. 
The units have been left in meters due to the wide range of possible wavebands
in use at CHARA.
\label{fig_UV}}
\end{figure}

A list of all available baselines is 
given in Table \ref{table_baselines}, and Figure \ref{fig_UV} shows the UV
coverage for all six telescopes for three hours either side of transit for
declinations of $-15^{\circ}$, + $30^{\circ}$ and $+75^{\circ}$.

\section{Facilities Overview}

There are five primary components to the CHARA facilities on Mount Wilson: the
telescopes and telescope enclosures; the vacuum light pipes and their mounting 
and alignment structures; the central beam synthesis and delay line building; 
the control
room and office building, and a small workshop. Each of these will be
discussed in a separate section below, except for the control room and
office building which is of standard construction, and the workshop
which makes use of an existing building on the mountain that originally served 
as the delay line and beam combining laboratory for the Mark III
Interferometer  \citep{MkIII}. 

\subsection{Light Collecting Telescopes}

Each of the six telescopes is a 1-m 
Mersenne-type a-focal beam reducer that injects a 
12.5-cm output beam into the vacuum transport tubes \citep{TR09}. 
The primary and secondary 
optics were manufactured as matched sets at LOMO in St. Petersburg, 
Russia, under a contract with Telescope Engineering Company of Golden, 
Colorado. The substrates for the 1-m primary and matched secondary 
are low-CTE Sitall 
and Zerodur, respectively. The typical primary mirror performance is 0.035 
waves rms at 633-nm and because the secondary and primary 
pairs are matched sets the combined performance of the telescope 
system exceeds this specification \citep{TR37}.  A seventh matched set 
has been fabricated for CHARA in St. Petersburg. These optics are an 
investment toward a future 
seventh telescope as well as a spare set should delays occur in periodic 
re-coating of existing optics.

\begin{figure}
\centerline{\includegraphics[scale=0.5]{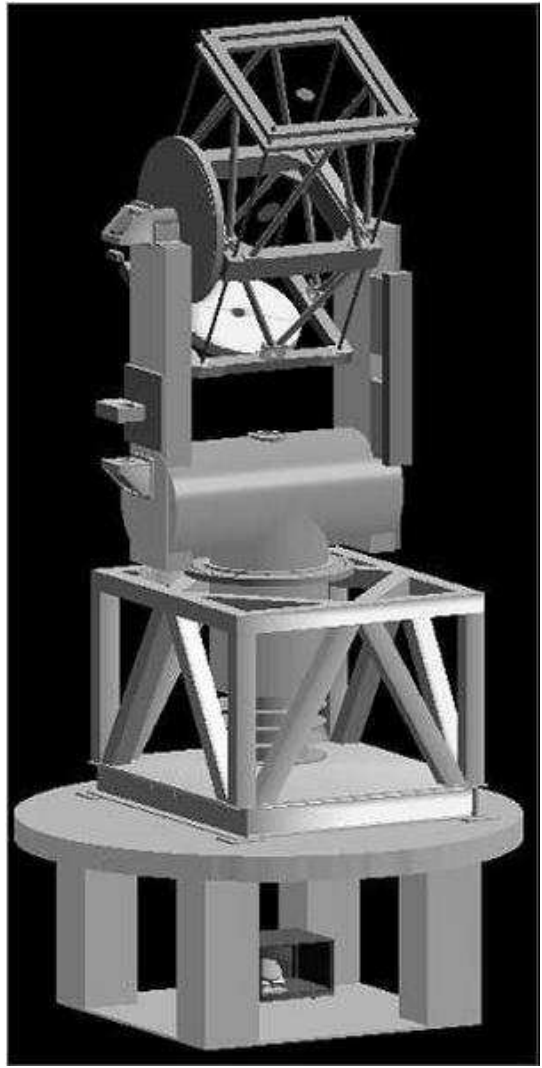}}
\caption{A CAD drawing of the CHARA telescope. 
The azimuth bearing is at the base of the fork, beneath the cable wrap
drums. Friction drive and idler rollers at the azimuth drive disk
complete the azimuth drive axis definition. The elevation bearings are
at the tops of the fork tubes, as is the elevation drive disk.
The entire telescope assembly rests on a concrete pier whose height
varies from telescope to telescope in order to provide the correct
location above the local terrain.  Beneath the telescope
lies the coud\'{e} box containing mirror \#7.
\label{fig_scope1}}
\end{figure}

\begin{figure}
\centerline{\includegraphics[scale=0.5]{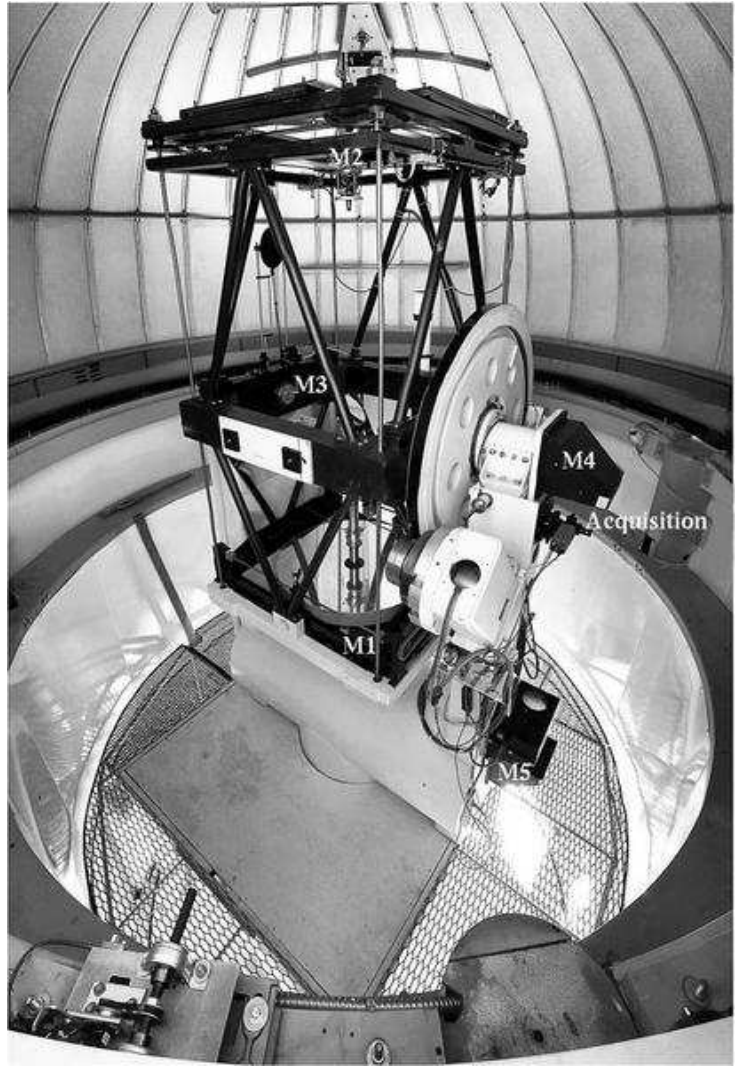}}
\caption{A composite picture of the W1 telescope. The elevation drive
system is on the right-hand side of the telescope. To the right of this
are the M4 and M5 mounts with the acquisition camera mounted
in between. The primary mirror M1 at the bottom of the telescope and 
the secondary M2 at the top are also
visible, as well as the small corner cube in the center of M2. M3 lies in
the center with M1 below it. Photograph by Steve Golden.
\label{fig_scope2}}
\end{figure}

CHARA commissioned a custom telescope mount design from Mr. Larry Barr.  
The design is a fork-style alt-azimuth mount with 7 mirror coud\'{e} beam
extraction. On one side of the fork, the telescope beam is available over
several meters in collimated space for acquisition, or future adaptive
optics correction. The mounts are exceptionally stiff and 
massive (23,000 pounds) for 
interferometric stability, and incorporate an Invar metering truss for 
temperature insensitive focus. This results in very high stability of
the focus of the telescopes which stay in alignment for up to a year
without the need for focus adjustment. The primary mirror cell utilizes a 
central radial support and 18-point whiffle tree, while the secondary 
support is actuated for adaptive tip/tilt control \citep{TR39}.  
M3 Engineering and 
Technology Corporation, of Tucson, fabricated and assembled the mounts. 
A CAD model of the telescope design is given in Figure \ref{fig_scope1},
while Figure \ref{fig_scope2} shows an actual telescope. 
\citet{Tel} and \citet{SPIE_Tel} give more
detailed descriptions of the CHARA telescopes.

Telescope pointing and tracking are controlled using ComSoft
TCS\footnote{www.comsoft-telescope.com} supplemented by 
Tpoints pointing model software\footnote{www.tpsoft.demon.co.uk}
to provide typical pointing accuracies of 20
arcseconds rms. A CCD camera located at the Nasmyth port provides for 
alignment to the central laboratory as well as for the initial acquisition 
of stars \citep{TR66}. The secondary mirror has a small corner cube 
in the center so
that an alignment laser sent out from the laboratory can also be seen in
the acquisition system. This ensures that the telescope
pointing and laboratory alignment match. Each telescope also has a
wide-field finder telescope and several small surveillance cameras to 
monitor telescope and enclosure status.

The telescope drives are DC servo motors supplied by 
Parker/Compumotor\footnote{www.parker.com}. 
These were selected in the hope that they would 
provide very smooth motion and still behave like stepper motors from a 
control point of view. In the early commissioning of the telescopes the 
drives were found to have a small, sub-hertz oscillation, which we 
had originally thought to be caused by the mechanical design. After a re-design 
of these drives we found that the oscillation was still there. It was 
later discovered to be caused by poor tuning of the servo system for the 
drive motors, and we reverted to the original mechanical design. 

Attached to the telescopes themselves is a custom CHARA designed and built 
telescope control system called the
``TElescope MAnager'' (TEMA). TEMA controls all the various telescope covers,
cameras, and alignment jigs in the dome as well as providing a local interface 
to the telescope control software. A local control pad and hand paddle allow 
an operator to control the telescope and dome
from within the enclosure, as well as remotely adjust mirrors in the
beam synthesis laboratory to align the optical delay beam axis to the
telescope beam axis.

\begin{figure}
\centerline{\includegraphics[scale=0.8]{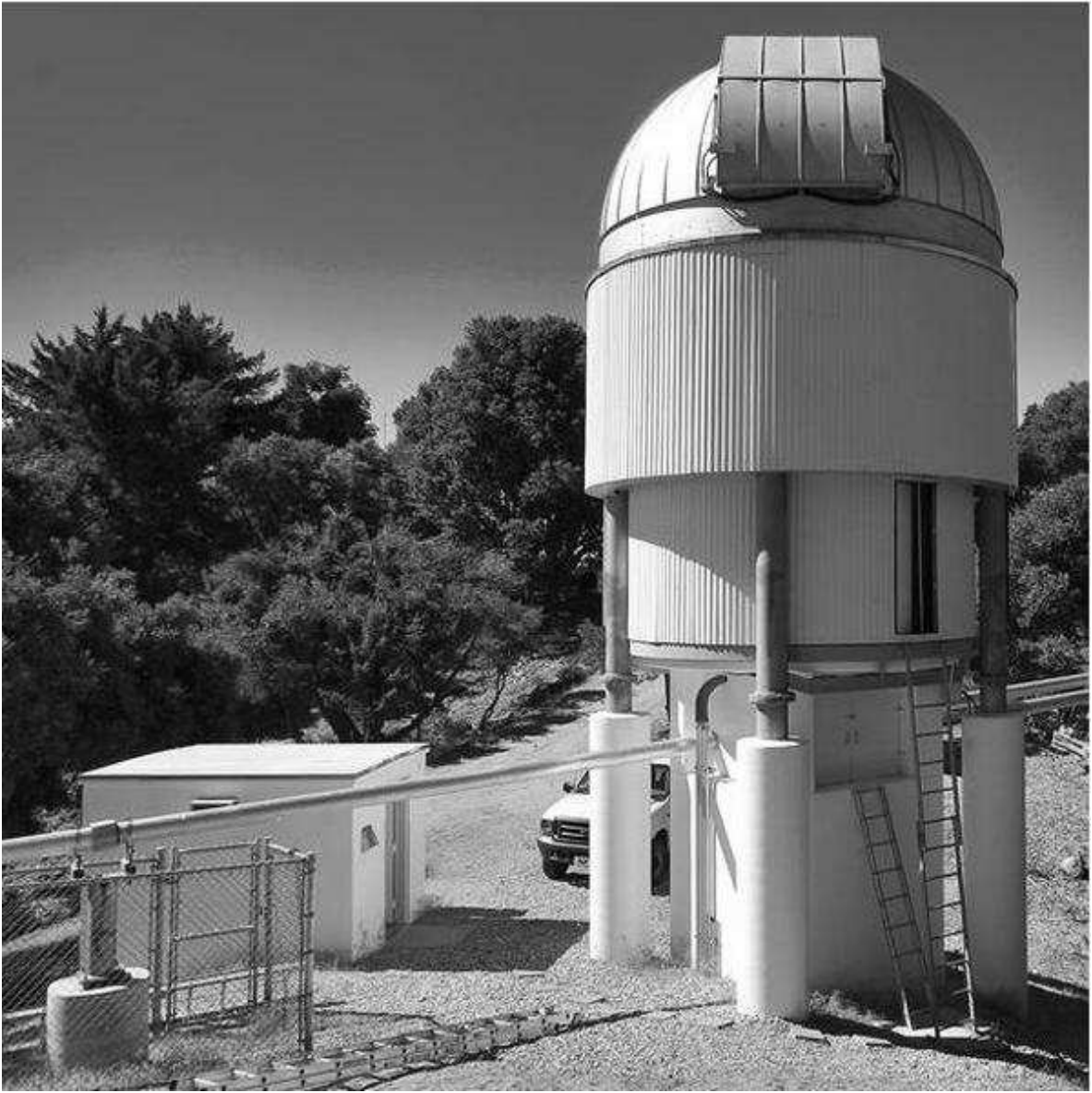}}
\caption{The E2 telescope enclosure and electronics bunker. The small 
entry door can be seen on the right of the lower cylinder. Both the upper 
and lower cylinders can move to allow a free flow of air around the 
telescope. The light pipe from E1 comes from the left, passes through the 
coud\`{e} area below the lower cylinder and leaves to the right with the 
light pipe from E2. Photograph by Steve Golden.
\label{fig_E2}}
\end{figure}

The telescopes are mounted on massive pedestals with a coud\'{e} area below
where the extracted beam is polarization compensated and directed to the
central laboratory. The surrounding enclosures are structurally
independent of the telescope pedestals, with lower and upper 
rings supporting a pair of
nested cylinder walls which open top and bottom for access and
ventilation. The walls provide $360^{\circ}$ free flow of air, which is
successful in suppressing dome seeing in spite of the very low wind speed
conditions typical at Mount Wilson.
The telescope enclosures incorporate 16.5-ft domes (provided by Ash 
Manufacturing with extra wide slits) slaved to the telescope azimuth.
CHARA's prime contractor on Mt. Wilson, 
Sea West Corporation\footnote{http://www.seawestinc.com}, 
of San Dimas, California, suggested this novel concentric cylinder design.
All enclosure
and telescope functions are remotely operable \citep{Enclgui}, 
and all telescope and enclosure control electronics, except TEMA are located 
in adjacent cement block bunkers to avoid heat production within the enclosures.
Figure \ref{fig_E2} shows the enclosure and electronics bunker for the E2
telescope.

All six telescopes are in place with all telescope optics installed and 
aligned, and all telescopes are in regular interferometric operation.
It is not necessary to visit the telescopes themselves under 
normal conditions to 
align, operate and close down the Array. The last telescope to be 
commissioned in interferometric mode
came on line in November 2003, and subsequently the Array has been
observing nightly.

\subsection{Vacuum Light Beam Transport Tube}

\begin{figure}
\centerline{\includegraphics[scale=0.8]{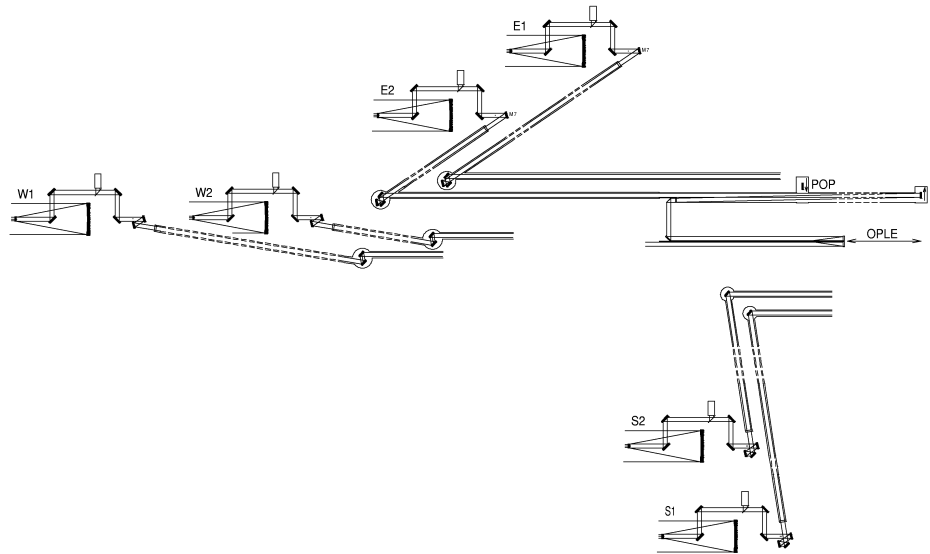}}
\caption{Schematic of the optical train bringing the light from each telescope
to inside the beam synthesis facility. Full optical symmetry has been 
maintained despite the three-dimensional nature of the mountain layout.
\label{fig_optics}}
\end{figure}

Seven reflections are required to direct the output beam from each telescope 
to the central beam combination facility, with mirror \#7 housed in the 
coud\'{e} box beneath the telescope. In order to preserve polarization symmetry 
around the three arms of the Y-array, one or two additional mirrors, 
depending upon the Array arm, are housed in this box. Figure \ref{fig_optics}
shows the layout of these relay mirrors. These extra reflections are
required because of the three-dimensional nature of the layout on the
mountain. In this way optical symmetry is maintained in all arms of the
interferometer \citep{Pol, TR28}. We have found no evidence for anomalous 
polarization effects in the data on the sky. The coud\'{e} optics 
inject the light through an optical window into a vacuum tube for 
transport to the Array center. Aluminum tubing with an outside diameter of 
20-cm accommodates the 12.5-cm light beams. Neoprene sleeves join thirty-foot 
lengths of this tubing, a technique that was first used by Michelson in his 
speed of light experiments on Mount Wilson and is also effectively 
used at other interferometers. 

Each telescope has a dedicated vacuum tube \citep{LightPipes}
so that each arm of the CHARA Array 
feeds two parallel tubes to the center. The terrain is such that the 
tubes must be elevated in places 
by up to 10-m above local ground level. Vertical 
support posts accommodate alignment and thermal expansion. Vacuum operation is 
not required for interferometric measurements in the near infra-red bands,
but does provide better image quality, especially in the early part of 
the night. Since mid-2004, all night-time operation takes place with the
system under vacuum. The vacuum pump is vibration isolated and it is 
possible to run the pump while observing and keep the vacuum at 0.5-Torr or
lower throughout the night.

\subsection{Central Beam Synthesis and Delay Line Building}

The delay lines, beam management subsystems, and beam combination subsystems 
are all housed in the central ``beam synthesis facility'' (BSF), an L-shaped 
building whose 94-m longest dimension houses the delay lines while the 
short part of the ``L'' comprises the ``beam combination laboratory'' (BCL). 
The BSF is a ``building within a building'' where the envelope between the 
inner and outer buildings serves as a heating and air conditioning plenum. 
This keeps the 1,000-m$^2$ of laboratory space at a reasonably constant and 
uniform temperature. The foundations and footings of the inner and outer 
structures are not structurally connected to minimize vibration transmission, 
and 24-inch thick concrete slabs beneath the optical tables and rail supports
provide stability to interferometric tolerances for all optical components.

\subsubsection{Optical Path Length Compensation}

Maintenance of zero optical path length difference is a major overhead for 
an optical interferometer with baselines of hundreds of meters. This 
is accomplished at the CHARA Array in two stages in an over-and-under 
arrangement. The first stage occurs while still in vacuum with six parallel 
tube systems (referred to as the ``Pipes of Pan'' or PoP's) \citep{TR04}
having assemblies 
with fixed delay intervals of 0-m, 36.6-m, 73.2-m, 109.7-m and 143.1-m,
that feed a mirror into the beam to reflect it back toward a 
periscope system that brings the beam into the continuous part of the delay
system. Incoming beams from the three arms of the Array are fed into the 
parallel PoP tubes by mirrors in large ``turning boxes'' that also serve 
to complete the polarization symmetry requirement. Down each PoP line are 
mirrors that can be moved into the beam to add, or remove, fixed amounts 
of delay.  At the end of each PoP line is a stationary ``end'' mirror that
represents the longest possible static delay available in the system.
Upon exiting the PoP's, 
beams leave vacuum and are injected into the continuous delay lines by a 
pair of periscope mirrors.

The continuously variable delay is provided by the 
``optical path length equalizers'' (OPLE's), a mid-generation design 
provided to CHARA under a contract with JPL in the evolutionary chain 
from the Mark III Interferometer to the Keck Interferometer  \citep{KECK}. 
The OPLE's are not in the vacuum system and
incorporate a cats eye arrangement in which an incoming beam 
reflects off one side of a parabola, comes to focus on a small 
secondary, returns to 
the other side of the parabola and then, re-collimated, is fed back parallel to 
the incoming beam.  Each cart rides on precision-aligned cylindrical steel 
rail pairs 46-m in length. A four-tiered nested servo system, with 
feedback from a laser metrology unit, provides 92-m of path length compensation 
tracking with an rms error of better than 20-nm, 
and typically as good as 10-nm. All 24 PoPs, six end mirrors, and all six 
OPLE carts are installed and are fully operational. 

\subsubsection{Beam Management}

Following path length compensation in the PoP/OPLE subsystems, the emerging 
12.5-cm beams of light are reduced to a final diameter of 1.9-cm using a 
two-element ``beam-reducing telescope'' (BRT). The reduced beams then pass 
through the ``longitudinal dispersion corrector'' (LDC) subsystem that 
corrects for the difference in air paths between two beams arising from the 
fact that the telescopes are not at the same elevation and that
part of the optical path is not in vacuum \citep{LDC1, LDC2}. 
The ``beam sampling system'' (BSS) provides 
the final stage of beam management. The BSS uses a dichroic beam-splitter 
and a flat mirror to separate visible from infrared light at the 1-$\mu$m 
boundary and then turns the two beams through 90$^{\circ}$ and sends them 
along parallel paths to the visible and IR beam combiners. 
Each BSS assembly is movable on a 
precision stage so that it also serves as a switch yard for selecting baseline 
pairs to be directed to the beam combiners. 

\begin{figure}
\centerline{\includegraphics[scale=0.9]{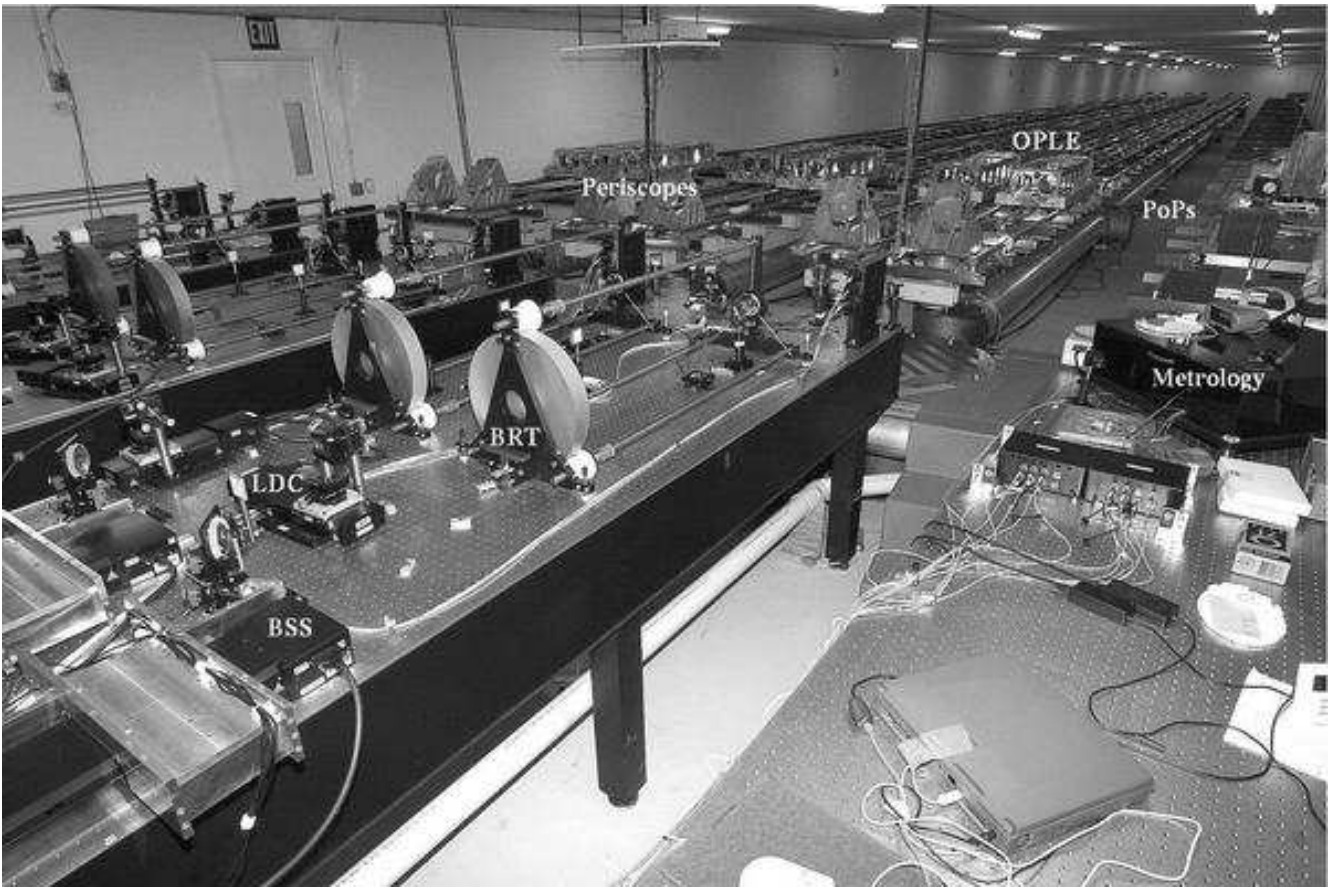}}
\caption{The delay line and beam management area. The PoPs, 
Periscopes, OPLEs, BRTs, LDCs and BSS system are clearly visible. Photograph by Steve Golden.
\label{fig_OPLE}}
\end{figure}

As with most subsystems in the Array, we are continually working to improve 
the beam management system. Nevertheless, all
six beam trains are complete and operational and all telescopes and
beam trains have been used for interferometry and are available for science. 
Figure \ref{fig_OPLE} shows a picture of the delay line and beam management 
area.

\subsubsection{Beam Combination}

The BCL currently houses five optical tables based upon functionality of 
subsystems. These tables contain the visible beam combiner, the infrared beam 
combiner, a HeNe laser for bore sighting/alignment and a white-light 
source for internal fringe generation, detectors for fringe and tip/tilt 
tracking, and a fiber-based beam combiner  \citep{FLUOR1, FLUOR2}
resulting from a collaboration with the Paris Observatory. 

At the time of writing,
all beam combination in the CHARA Array has been restricted to
two-beam systems. Future upgrades in 2005 will include a three-way open-air beam
combiner capable of measuring three baselines and a
single closure phase. The tests are already underway of a four- to 
six-way fiber-based
beam combiner being designed and constructed by Dr. John Monnier at the
University of Michigan  \citep{MIRC}. 
In late 2004, a new grant from the National
Science Foundation was awarded to Dr. Monnier and collaborators at
Georgia State University and the Michelson Science Center to fund a six-way 
IR-band fringe tracker. We expect the detailed design and
construction of this system to begin in 2005.

It was decided early on in the design phase of the CHARA Array to separate the
functions of fringe tracking and fringe amplitude and phase 
measurement  \citep{BC}. In this way, each system can be optimized for one 
task or another
and development of the various beam combiners can go on in parallel. We
are exploring several methods of fringe tracking, including packet
tracking where one keeps the fringe packet centered in a long scan, group
delay tracking where one keeps the center of the group delay in position,
and phase locking where one tracks on a single fringe. It is expected
that all three methods will be used under different conditions and for
different science targets. For example, it will be possible to phase
lock using the IR system and send a phased beam into the visible
beam combiner. The reverse will, of course, also be possible. It will even be
possible to divide the six telescopes between several beam combiners and
thereby create separate interferometers working in parallel.

Two high-speed CCDs have been purchased from Astronomical Research 
Cameras of San Diego for use in the visible beam combination area, one
for the I and R band beam combiner and one for tip/tilt detection.
We have also acquired a fiber coupling stage that breaks the full aperture into
seven smaller apertures and couples each of these to a multi-mode fiber. 
The central beam includes the telescope 
secondary obscuration and is therefore used 
primarily for alignment. This allows us to have seven small (30-cm on the sky) 
aperture systems feeding a fiber-based 
slit in the low resolution spectrograph and process all seven
systems in parallel  \citep{VIS}. The construction
of this visible light system is in progress, and we expect to 
have achieved first fringe on the sky by late 2004.

A higher spectral resolution spectrograph is also under construction and 
will use the same sub-apertures and fibers \citep{Spec}. 
This spectrograph is intended for combined high spatial and high spectral
resolution measurements.
The spectrograph is a modified Ebert-Newtonian based on the design used 
for the Georgia State Multi-Telescope Telescope \citep{MTT}. 
It is fiber-fed and will produce coherent, spatially resolved spectra. The 
use of this spectrograph on the sky is delayed until phase locking is possible
at the CHARA Array so that phased beams can be sent into the system. This 
should be possible in late 2005.

The second camera is being tested as a tip/tilt detection upgrade. 
In lieu of a full-up visible 
light beam combiner, the visible channel short of 600-nm is presently 
used entirely for photomultiplier based tip/tilt detection for which the 
high bandwidth limiting magnitude is V = +9.5 at 10-ms 
sample times, although in 
good seeing conditions that permit the use of longer integration times, we 
can track objects as faint as V = +12.0. 
This tip/tilt detection scheme and correction algorithm is largely based on 
the system developed at the Sydney University Stellar Interferometer (SUSI) 
\citep{susi_tiptilt}. Light from 
600-nm to 1-$\mu$m is divided between an intensified CCD, allowing for image 
quality inspection and alignment, and the low-resolution spectrograph used 
for I and R band interferometry. The bore sight laser and white 
light sources are 
regularly used for the essential tasks of alignment and overall system 
fine-tuning. The ICCD also supplies us with a reference 
position defining the optical axis of the interferometer, and we regularly
align the tip/tilt detectors to ensure that star images from each telescope
lie on this optical axis.

The layout of the current IR beam combiner, dubbed ``CHARA Classic", 
is that of a simple pupil-plane beam combiner  \citep{IRBC}. The two outputs 
from the beam splitter are separately imaged onto two spots on the beam 
combiner camera (described below), with fringes detected in a scanning 
mode provided by dithering a mirror 
mounted to a piezoelectric translation stage. The stage is driven with 
a symmetric saw tooth signal whose response to a given driving signal 
was mapped with a laser interferometer to provide a 1-kHz data acquisition 
rate corresponding to five samples per fringe at a typical 200 Hz fringe 
frequency. Sample rates of 250, 500 and 750 Hz are also possible.
The current K limiting magnitude for fringe detection with this 
system is K = +6.5 for raw visibilities of 0.50. 

This system is also capable of IR phase locking, which has successfully been 
achieved in the lab, with the goal of the system working on starlight
in 2005. This will improve throughput from one sample every 0.5 sec to 80 
samples/sec and ultimately provide the ability to fringe track in H band 
with CHARA Classic and gain several magnitudes in K band using the fiber 
based beam combiner described below. It will also 
be possible to phase lock in one of the IR bands in order to provide 
phased beams for both the low- and high-resolution visible
spectrographs.

In our current fringe scanning 
method the maximum sample time is 4-ms, however, with the new phase 
locking method it is possible to increase this to 20-ms or more.
Furthermore, all mirror surfaces are currently coated with Al, and the 
remaining project funding provides for upgrading a number of relay mirrors 
in vacuum or in sheltered environments with over-coated silver. Our
calculations show \citep{TR91} that silver coatings will double the near IR
throughput and the expected gain in the visible red is much larger.
These and other steps are expected to increase sensitivity to K = +9.0. 

Through a collaboration with the Paris Observatory, a second beam 
combiner has been installed at the CHARA Array. The Fiber-Linked Unit for 
Optical Recombination (FLUOR) has been in use at the IOTA interferometer 
on Mt. Hopkins since 1995 (e.g. \citet{Perrinetal}). The special feature of 
FLUOR is that it incorporates optical fibers as spatial filters and a 
fiber X junction as the interfering element instead of a beam splitter. The 
combination of spatial filtering with subsequent photometric monitoring
of beam intensities permits this instrument to measure visibilities 
of bright objects with precisions of better than 1\%. FLUOR is
a key tool for problems such as accurate measurement of variable star 
pulsation, stellar shapes for main sequence and pre-main sequence stars, 
large $\Delta$m binaries, and limb darkening determination through measurements 
beyond the first null in visibility that require such high accuracies.
The FLUOR beam combiner has been used regularly on the sky since early
2003.

\subsubsection{IR Beam combiner camera}

The heart of the infrared beam combiner is a Rockwell 
256$\times$256 HgCdTe PICNIC 
array with 40-$\mu$m pixels possessing QE's better 
than 60\% in the 1 to 2.3-$\mu$m 
spectral regime. The CHARA ``Near Infrared Observer'' (NIRO) camera was 
built at Georgia State with the kind assistance of Drs. Wes Traub and 
Rafael Millan-Gabet of the Harvard-Smithsonian Center for Astrophysics. 
NIRO is thus very similar to the camera used at the IOTA 
interferometer \citep{MSTC1999, NIRO}. 

The PICNIC array is sensitive to light at wavelengths of 0.8 to 2.5-$\mu$m and 
is electrically structured as four 128$\times$128 quadrants, each with its own 
output amplifier; NIRO is wired to read out only one quadrant. Array 
control and data acquisition are controlled by a digital I/O board
in a PC and a custom-built interface box containing a 
clock driver board that provides the DC bias levels for the array 
and an analog-to-digital converter (ADC) board. The input to the ADC board 
is the array's analog output signal, buffered by a low-noise 
20$\times$ pre-amplifier 
mounted on the inside of the dewar lid (operating near room temperature). 
The ADC is a 16-bit unit capable of 100 kilo-samples per second.

The PICNIC array is mounted in an Infrared Laboratories HDL-5 dewar modified 
for use with liquid nitrogen in both cryogen cans. Vacuum in the dewar is 
maintained to better than 10$^{-8}$-Torr by a 
combination of molecular sieves and 
an ion pump, resulting in a dewar hold-time of 22 
hours. A 60$^{\circ}$ off-axis parabolic mirror focuses the two collimated 
interferometer beams into two spots on the array after passing through a 
fused silica vacuum window and a cooled filter. 
The filter wheel can be rotated manually and holds up to six 
filters. Currently, it holds H and K$^{\prime}$-band filters, as well 
as several narrow-band filters within the K band.

\section{Control System}

The Array control system is based upon real-time Linux except for the 
OPLE subsystem, which was developed by JPL and runs under VxWorks, 
and the FLUOR beam combiner, which utilizes Labview. The control system is 
fully operational although it is subject to frequent updates and improvements 
as observing experience is gained. 

The control model consists of several layers. At the lowest level is the
real-time code which is slaved to a 1-ms clock tick derived
from GPS and distributed around the entire Array. In this way we can run 
synchronous tasks across many CPUs distributed throughout the facility.
In the second layer, each real-time process
is controlled by a server program run in the Linux environment. This
server provides a command line interface that provides full
control of the system in question as well as many engineering test
routines and functions. Each server can be logged into remotely via a
simple text-base interface for remote engineering and test operations.
At the highest level, more user-friendly GUI programs based on the GTK
windowing system can connect via a TCP/IP socket to any server and provide 
all the functionality normally required for observing. Sequencing is done by a
single client program, dubbed the ``Central Scrutinizer''\footnote{Named 
after an all controlling character in Frank Zappa's 
``Joe's Garage''}, that connects 
to all servers and controls the acquisition of stars, data collection and
writing a log of all activity. Acquiring a new target consists of
entering an HD, SAO or IRC number and clicking a single button.
The Central Scrutinizer also
automatically collects baseline solution and telescope pointing data.

Because of the client/server model used, the CHARA Array can be controlled 
from a single remote CPU via a secured internet connection and
this accessibility has been utilized 
to establish the Cleon C. Arrington Remote Operations Center (AROC), 
located in Atlanta on the Georgia State campus. 
A second remote control facility has been established on the grounds of
the Paris Observatory and was used to control the Array for the first
time in June of 2004. A remote capability for the FLUOR beam combiner was added
in late 2004 \citep{PROC}.

An IDL-based observing planning tool for CHARA has been developed \citep{TR90}
that provides a complete tool set for planning observations with the CHARA 
Array: providing UV coverage plots; times when a particular object 
is available for observation; theoretical visibility plots; and aids in 
the selection of telescope and PoP configurations. This planning tool can 
be downloaded at http://www.noao.edu/staff/aufdenberg/chara\_plan.

\section{Data Processing}

While we fully expect our views and methods on data analysis
to evolve with time we will discuss here the two methods currently in use
at the CHARA Array. Both are based on the work of \citet{Benson_A} and
yield results consistent with one another, though the second method is more 
reliable in low signal to noise data.

We write the fringe signal on each detector pixel as
\begin{equation}
D_i(t) = \frac{I_{i1} + I_{i2}}{2} + \, (-1)^i \,
                \sqrt{I_{i1} I_{i2}}
                \, \nu \, {\rm sinc}(\pi \, \Delta\sigma \, v_g t)
                \cos( 2 \pi \sigma_0 v_g t + \phi)
\end{equation}
where $I_{ij}$ is the light intensity that reaches detector $i$ from 
telescope $j$, $\nu = V \times V_{\rm sys}$ is the correlation of 
the two beams related
to the visibility of the object $V$ and the system visibility $V_{\rm sys}$,
the optical filters are centered at wave number
$\sigma_0$ and have a band pass of $\Delta\sigma$, $v_g$ is the
group velocity of the fringe packet resulting from the motion of
the dither mirror, and $\phi$ is the sum of the fringe phase and the
atmospherically induced phase error.

This must be calibrated for beam intensity, so this expression is
normalized using a low-pass filtered version of itself  resulting in
\begin{equation}
N_i(t) = 1  + \, (-1)^i \,
        \frac {2 \sqrt{I_{i1} I_{i2}}}{I_{i1} + I_{i2}}
                \, \nu \, {\rm sinc}(\pi \, \Delta\sigma \, v_g t)
                \cos( 2 \pi \sigma_0 v_g t + \phi)
\end{equation}
where instead of intensities we now have the normalized signal
$N_i(t)$ and we see that there will be a transfer function due to the
light intensities for each detector channel given by
\begin{equation}
T_{i} = \frac {2 \sqrt{I_{i1} I_{i2}}}{I_{i1} + I_{i2}}.
\end{equation}

\begin{figure}
\centerline{\includegraphics[scale=0.6]{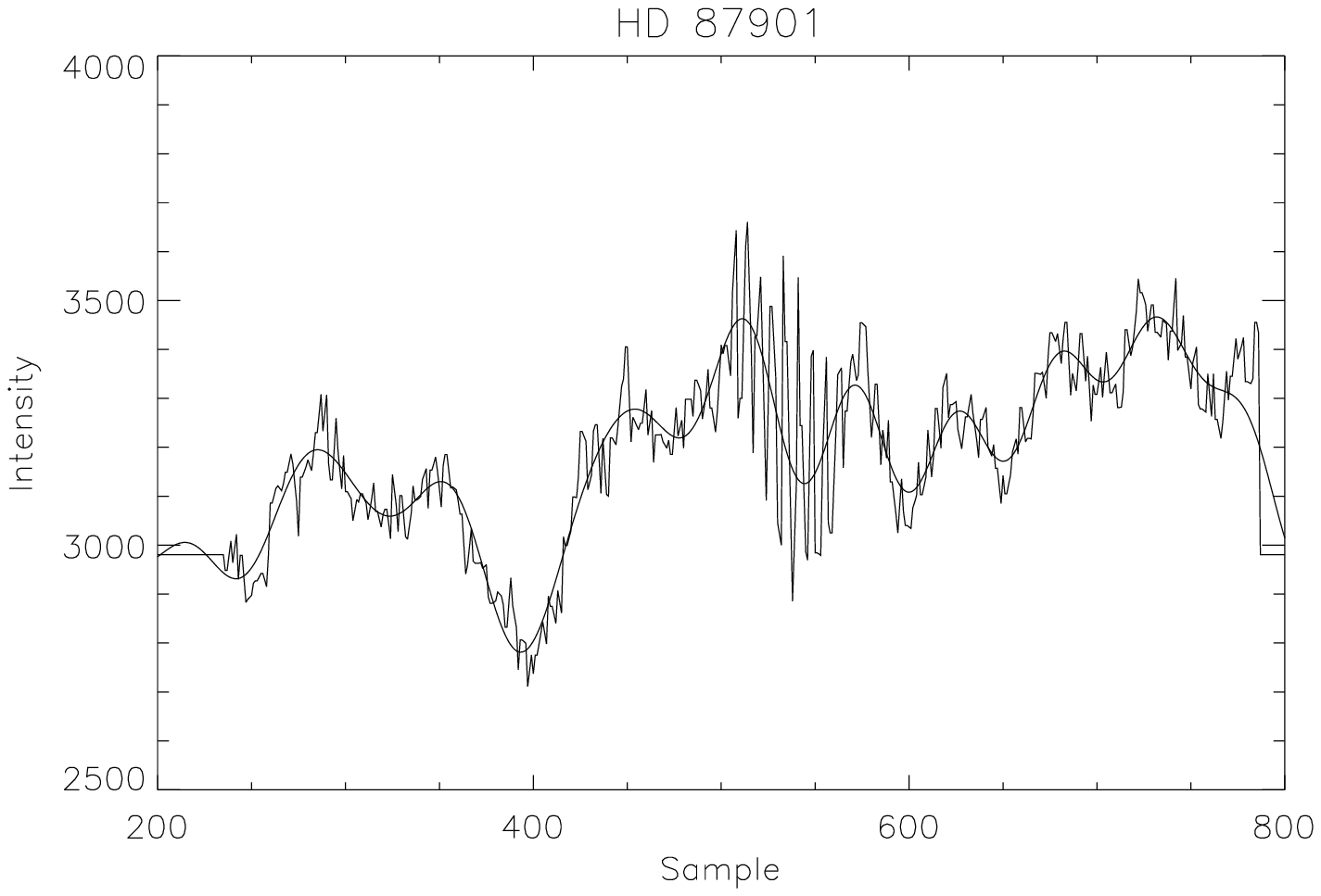}}
\vskip 0.1in
\centerline{\includegraphics[scale=0.6]{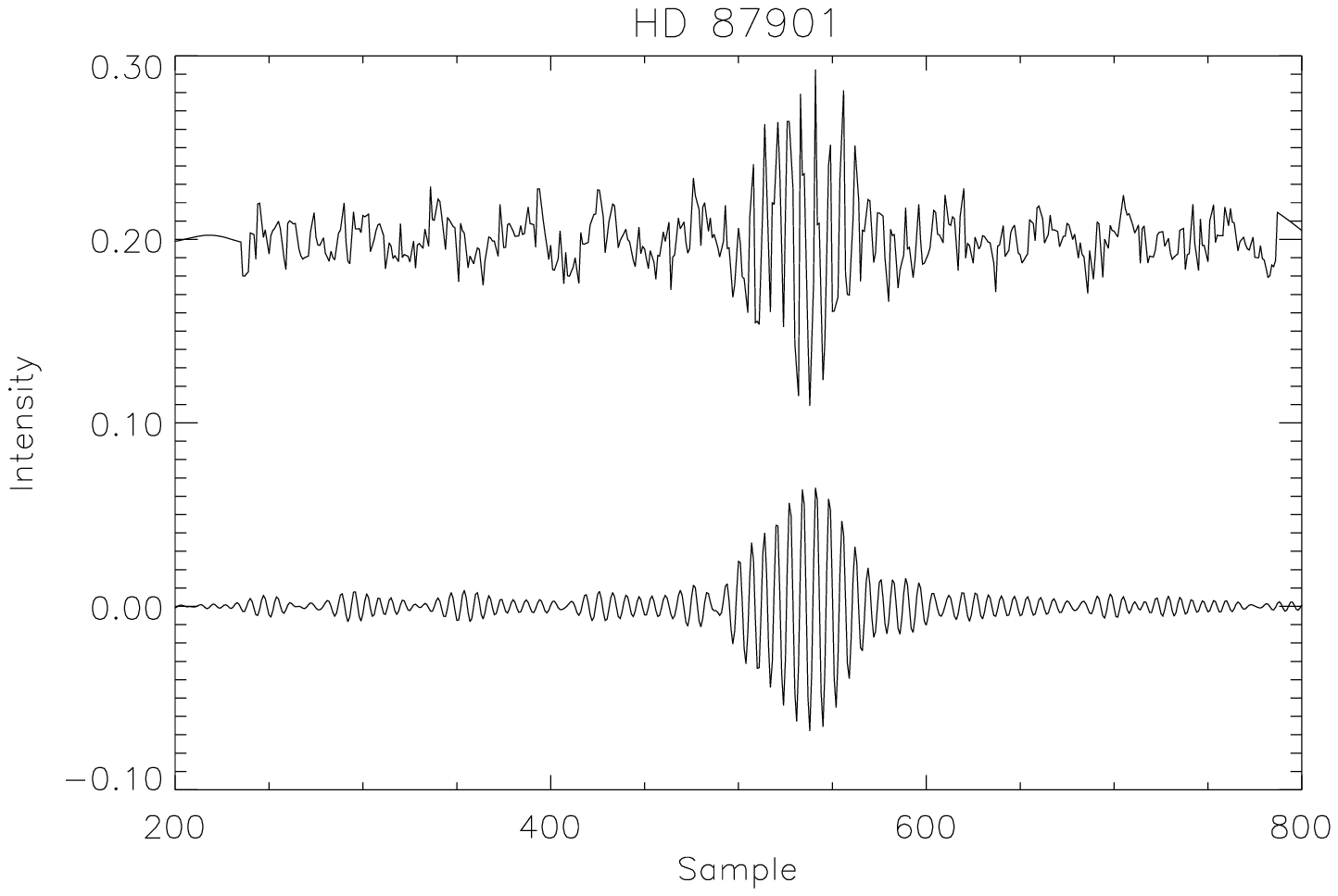}}
\caption{A fringe scan is shown above in its raw signal form (top) with the 
low-pass filtered version superimposed prior to normalization. The bottom 
frame shows the same scan, with an offset of 0.2 for clarity,
after normalization and, finally, after 
implementation of the band-pass filter.
\label{fig_fringe_example}}
\end{figure}

In the first data reduction method used at the CHARA Array, we put the 
normalized signal $N_i(t)$ through a bandpass 
filter centered at the frequency expected due to the dither mirror
motion. An example processed fringe, 
from the rapidly rotating star Regulus (HD 87901) 
observed on 16 April 2004 with 
the E1-S2 baseline, is shown in Figure \ref{fig_fringe_example}.  
The top frame shows the raw 
fringe with the smoothed version of that fringe, obtained from the low-pass 
filtering, superimposed. The middle frame (shifted by 0.2 in relative 
intensity for clarity) is that fringe normalized to the smoothed version, 
and the bottom frame is the result of bandpass filtering after which the 
maximum absolute excursion from zero relative intensity is a measure of 
$\nu$. These fringes exhibit symptoms of longitudinal dispersion, 
which will be corrected by the LDC system mentioned previously. For most
high signal to noise cases this analysis method works well, though it
has a tendency to overestimate the correlation for low signal to noise
data.

A second method of analysis is based on a spectral analysis. If we
compute the power spectrum of the fringe signal we get
\begin{equation}
{\rm PS}\left[ N_i(t) - 1 \right] = T_i^2
        \left[ \frac{\nu}{\Delta \sigma \, v_g} \right]^2
        \left[ \Pi \left(\frac{\nu - \sigma_0 v_g}{\Delta \sigma \,
v_g}\right)+        \Pi \left(\frac{\nu + \sigma_0 v_g}{\Delta \sigma \,
v_g} \right)\right]\end{equation}
where
\begin{equation}
\Pi(x) =
        \left\{
        \begin{array}{ll}
            0, &  |x| > \frac{1}{2}\\
            \frac{1}{2}, &  |x| = \frac{1}{2}\\
            1, & |x| < \frac{1}{2} \\
        \end{array}
        \right.
\end{equation}
and since the fringe signal is a real signal, we can safely ignore
the negative spatial frequencies and write
\begin{equation}
{\rm PS}\left[ N_i(t) - 1 \right] = T_i^2
        \left[ \frac{\nu}{\Delta \sigma \, v_g} \right]^2
        \Pi \left(\frac{\nu - \sigma_0 v_g}{\Delta \, \sigma v_g}
\right).
\end{equation}
The total integral of the power is then
\begin{equation}
S_i = \int d\nu \, {\rm PS} \left[ N_i(t) - 1 \right] =
                T_i^2 \frac{\nu^2}{\Delta\sigma \, v_g}
\label{eq_v2}
\end{equation}
and this results in an estimate of $\nu^2$.

\begin{figure}
\centerline{\includegraphics[scale=0.6]{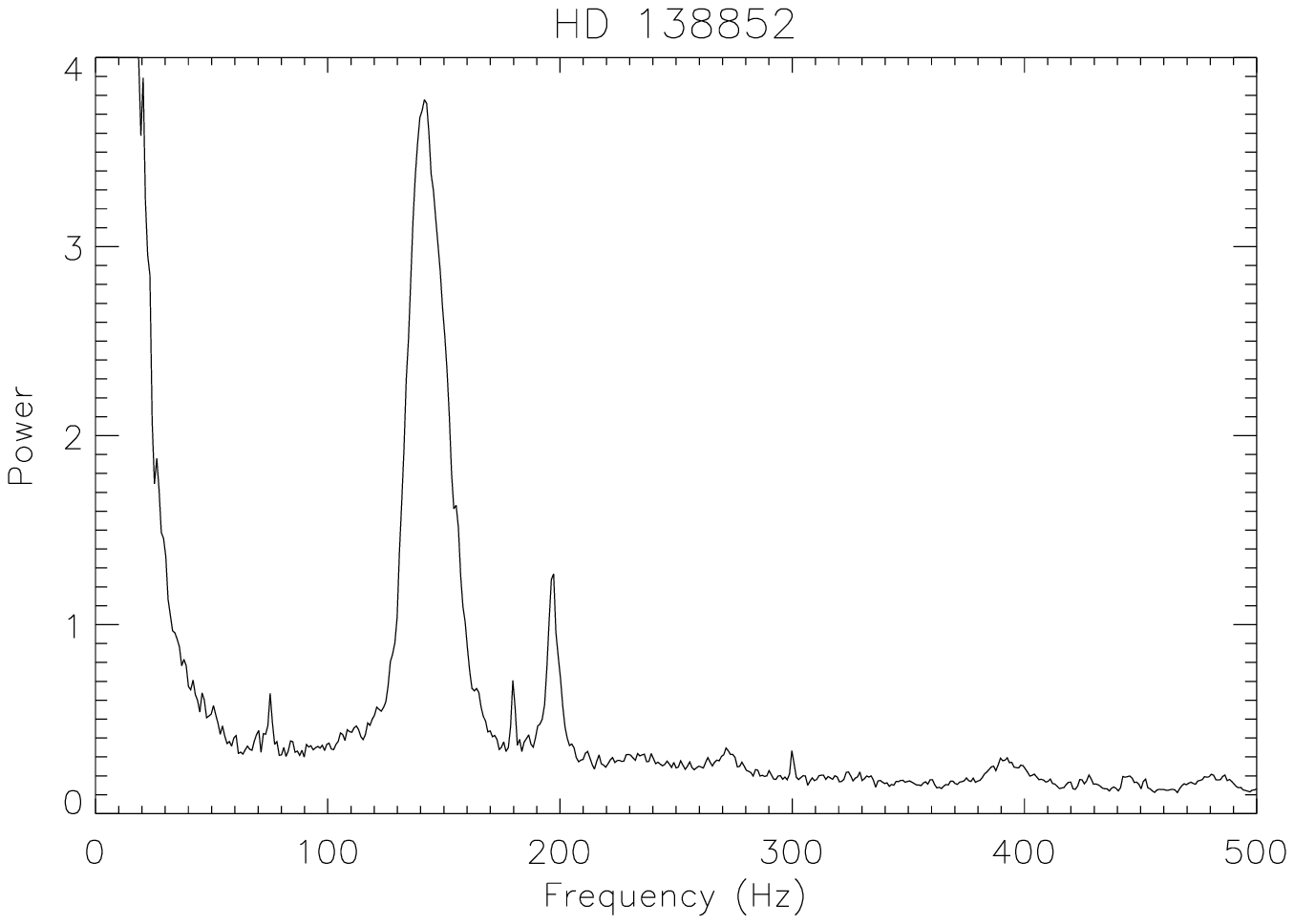}}
\vskip 0.1in
\centerline{\includegraphics[scale=0.6]{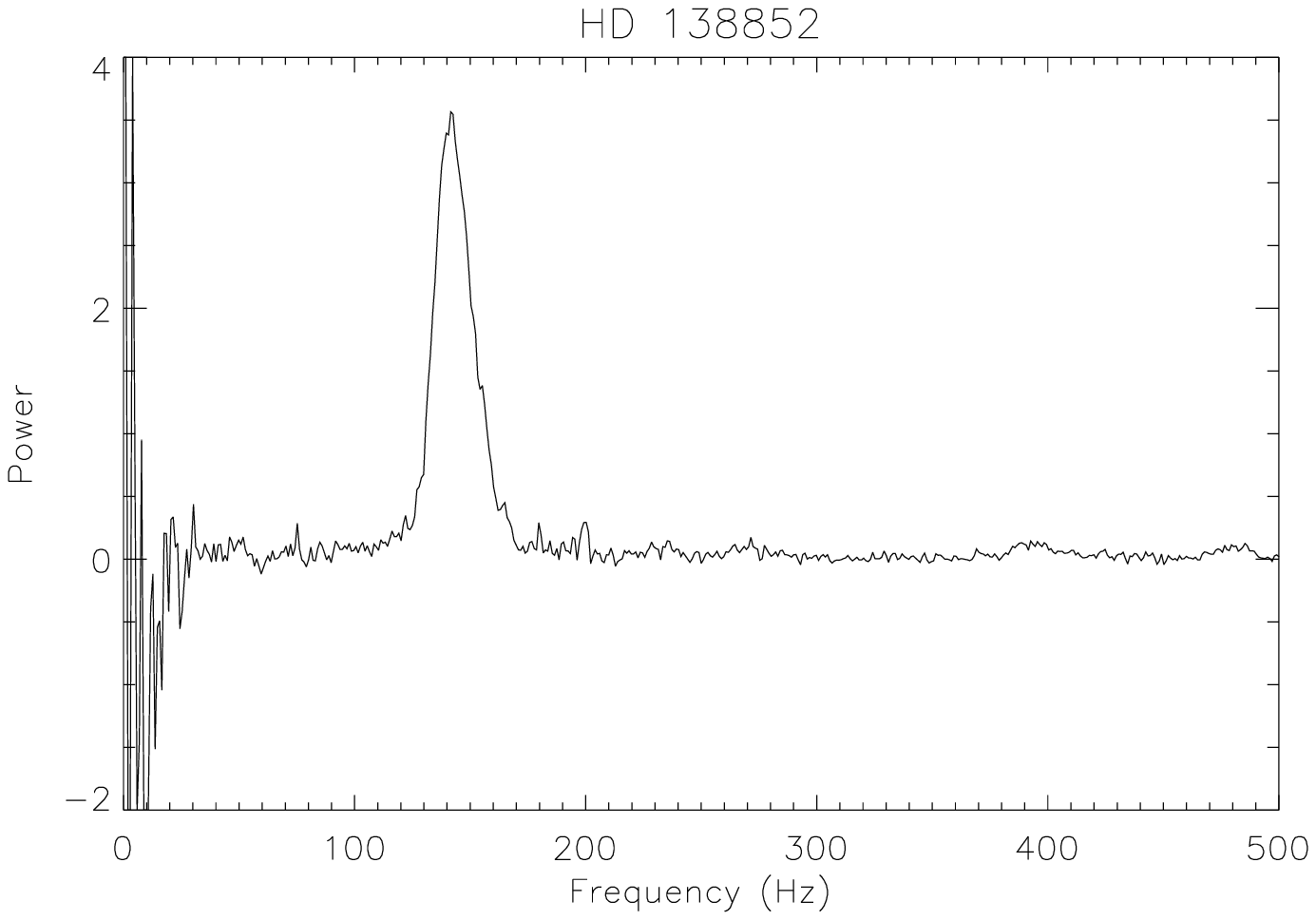}}
\caption{An example of a raw power spectrum (top) showing a combination
of the fringe signal, photon noise, scintillation noise and camera noise.
Once the noise has been measured and removed only the fringe signal remains 
(bottom).
\label{fig_psir}}
\end{figure}

Since there will also be detector, scintillation and photon noise in the
signal, these must be removed before the final integration is performed.
Each data set is followed by a series of data scans incorporating shutters 
to provide data from
each telescope separately and without light from either telescope. The
power spectrum with both shutters closed
contains only detector noise, while those with light from the telescopes
contain detector noise as well as half the scintillation and photon noise.
This contains enough information to remove the noise bias as well as
to estimate the transfer function $T_i$. An example power spectrum of a
fringe signal taken of the star HD 138852 on June 6th 2004
is given in Figure \ref{fig_psir}.

There is one more correction that can be made to the correlation estimate. 
Because of atmospheric turbulence, the correlation is changing 
constantly and so can be considered to be a random variable with some
mean $\overline{\nu}$ and a variance $\sigma_{\nu}^2$. Since we are
measuring the mean of the square of the correlation, we are actually
measuring
\begin{equation}
\overline{\nu^2} = \overline{\nu}^2 + \sigma_{\nu}^2,
\end{equation}
and thus all estimates of the square of the correlation are biased by the
variance of the correlation. Unfortunately, it is not possible to take
the square root of $S_i$ as, due to the statistical nature of the
measure, it is sometimes negative. It is, however, possible to square
$S_i$, resulting in an estimate of $\overline{\nu^4}$. If one assumes
the statistical distribution of the correlation is normal, one can then
form the unbiased estimator for the correlation
\begin{equation}
\overline{\nu} = \left(\frac{3\overline{\nu^2}^2 - \overline{\nu^4}}
			{2}\right)^\frac{1}{4}
\label{eq_norm}
\end{equation}
with the corresponding variance estimate
\begin{equation}
\sigma_{\nu}^2 = \sqrt{\overline{\nu}^4 - \frac{1}{2}(\overline{\nu^2}^2
			- \overline{\nu^4})} - \overline{\nu}^2.
\end{equation}

Since the real correlation can never be negative, it is sometimes better
to use a log-normal distribution, normally parametrized using the variables 
$\mu$ and $\sigma^2$. These can be determined using
\begin{equation}
\mu = \frac{1}{4}\ln\frac{\overline{\nu^2}^4}{\overline{\nu^4}}
\end{equation}
and
\begin{equation}
\sigma^2 = \frac{1}{4}\ln\frac{\overline{\nu^4}}{\overline{\nu^2}^2}
\end{equation}
from which we get the unbiased correlation estimate
\begin{equation}
\overline{\nu} = \exp(\mu + \frac{1}{2} \sigma^2)
\label{eq_lognorm}
\end{equation}
with the variance
\begin{equation}
\sigma_{\nu}^2 = \exp(2\mu + 2\sigma^2) - \exp(2\mu + \sigma^2).
\end{equation}

\begin{figure}
\centerline{\includegraphics[scale=0.8]{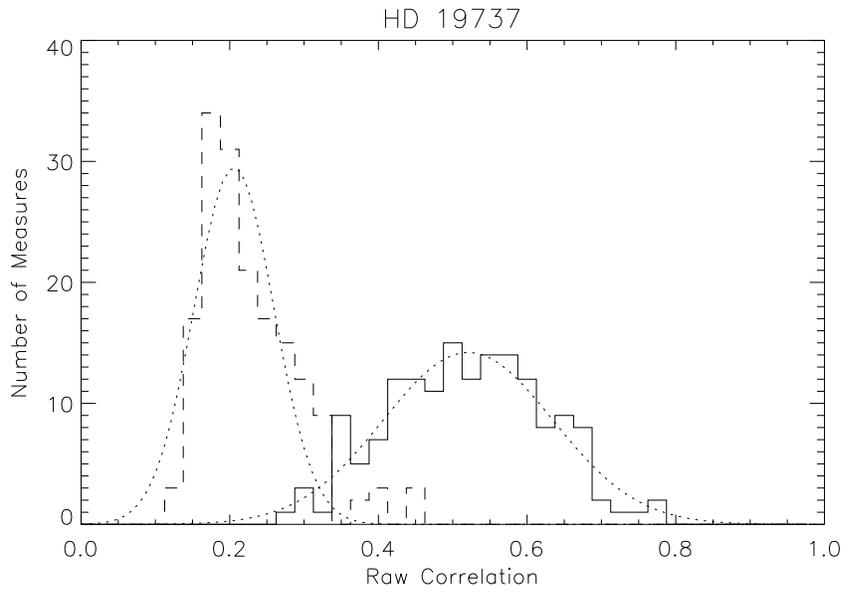}}
\caption{Two histograms of raw correlation as measured on the same object 
HD 197373 on two different nights, one of good seeing 2004-06-20 
(solid line) and one of poor seeing 2004-06-29 (dashed line). The actual
correlation at both times should be close to unity. The dotted lines show the 
best fit Gaussian curve for each data set.
\label{fig_vhist}}
\end{figure}

In practice the normal and log-normal equations give virtually the same 
results during times of good seeing, or high correlation, 
and typically fail at very low signal 
to noise. An example of some raw correlation data is given in Figure 
\ref{fig_vhist}. In this figure histograms of the raw correlation
measured during two separate measurement sequences are shown for the
same object, one taken during a time when the seeing was approximately 1
arcsecond and one when the seeing was approximately 2 arcseconds. 
Superimposed on these plots are the best fit Gaussian curves. In the
case of good seeing the histogram is very symmetric and well
approximated by a normal distribution, whereas the histogram made during poor
seeing is very asymmetric and a log-normal distribution is required.

The effect of the correlation bias discussed above is also quite clear
in these data. In the reduction of the data taken during good seeing
the correlation estimates based directly on equation (\ref{eq_v2}) is
$0.542\pm0.110$, where the error quoted here is the formal standard deviation
of the measurement, while those based on the de-biased estimates in
equation (\ref{eq_norm}) and (\ref{eq_lognorm}) are $0.542\pm0.185$ and
$0.543\pm0.100$ respectively. All three measures agree. On the other hand, these
estimators produced $0.180\pm0.091$, $0.145\pm0.165$ and $0.172\pm0.084$
in the poor seeing example, a difference of 8\%. Clearly, this measurment bias 
is very important unless the conditions and instrument performance are both 
very good.

This bias due to correlation variance is not significant in fiber based
beam combiners such as FLUOR due to the
spatial filtering of the beams, and will also be much less important in
smaller aperture interferometers.
Still, in the case of CHARA's larger apertures is 
certainly does affect the final 
calibrated result, and our investigation of this correlation bias 
process continues.

A typical dataset for a target star is a sequence of scans in which the 
target is interleaved between scans of a reference or calibrator star that 
has been selected on the basis of having a well-known visibility or an
expected visibility near unity. 
Time variations in instrumental visibility due to seeing and other effects 
are compensated for by interpolating the target epochs to the calibrator 
epochs and then dividing the measured correlations by the interpolated 
calibrator visibilities corrected for the expected calibrator visibility. 
Software has also been
written to enable us to use the output of this analysis and put it in
the correct format for use with the calibration and data reduction
software produced by the Michelson Science Center \citep{PTI_calib}.

\begin{figure}
\centerline{\includegraphics[scale=0.6]{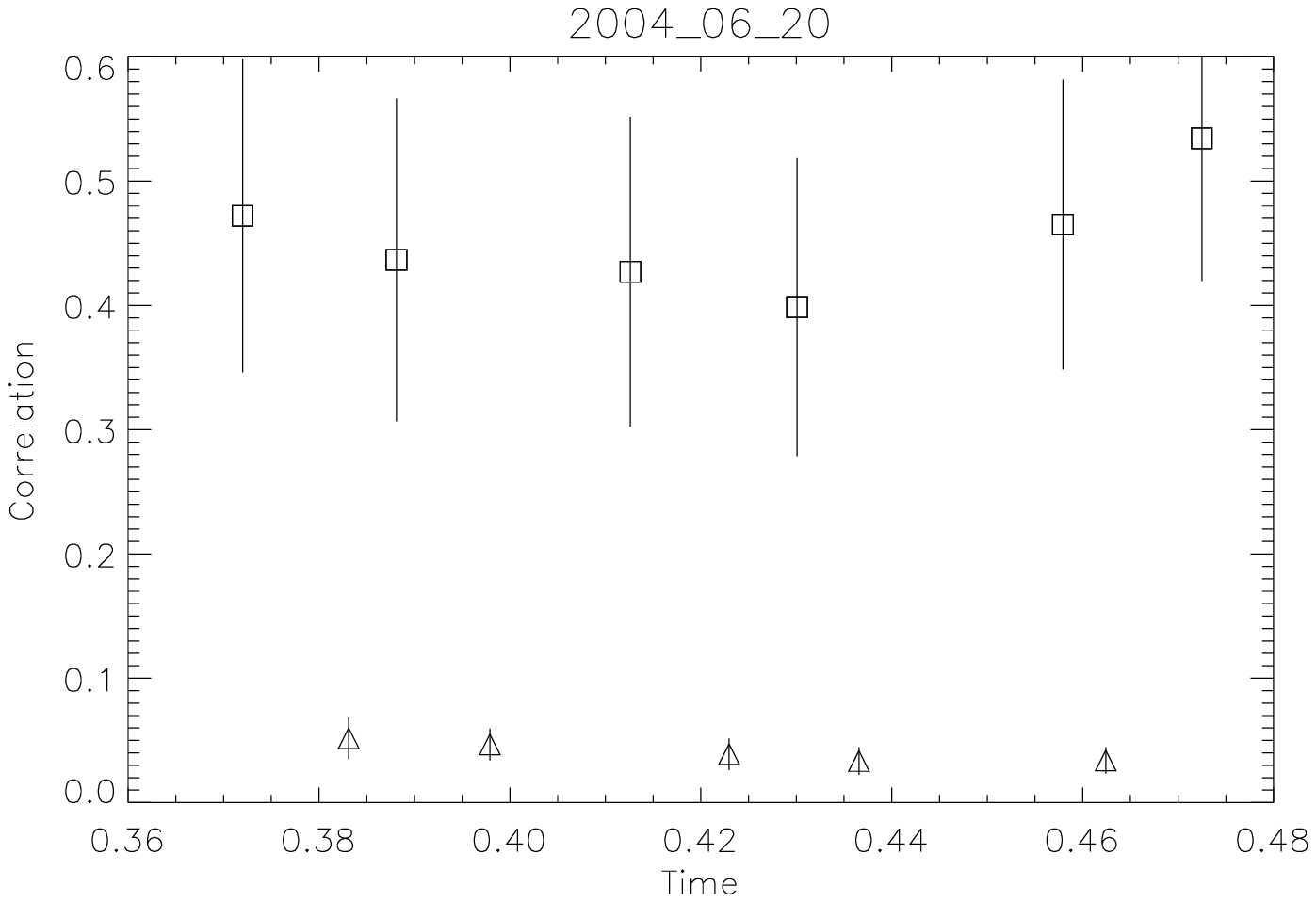}}
\vskip 0.1in
\centerline{\includegraphics[scale=0.6]{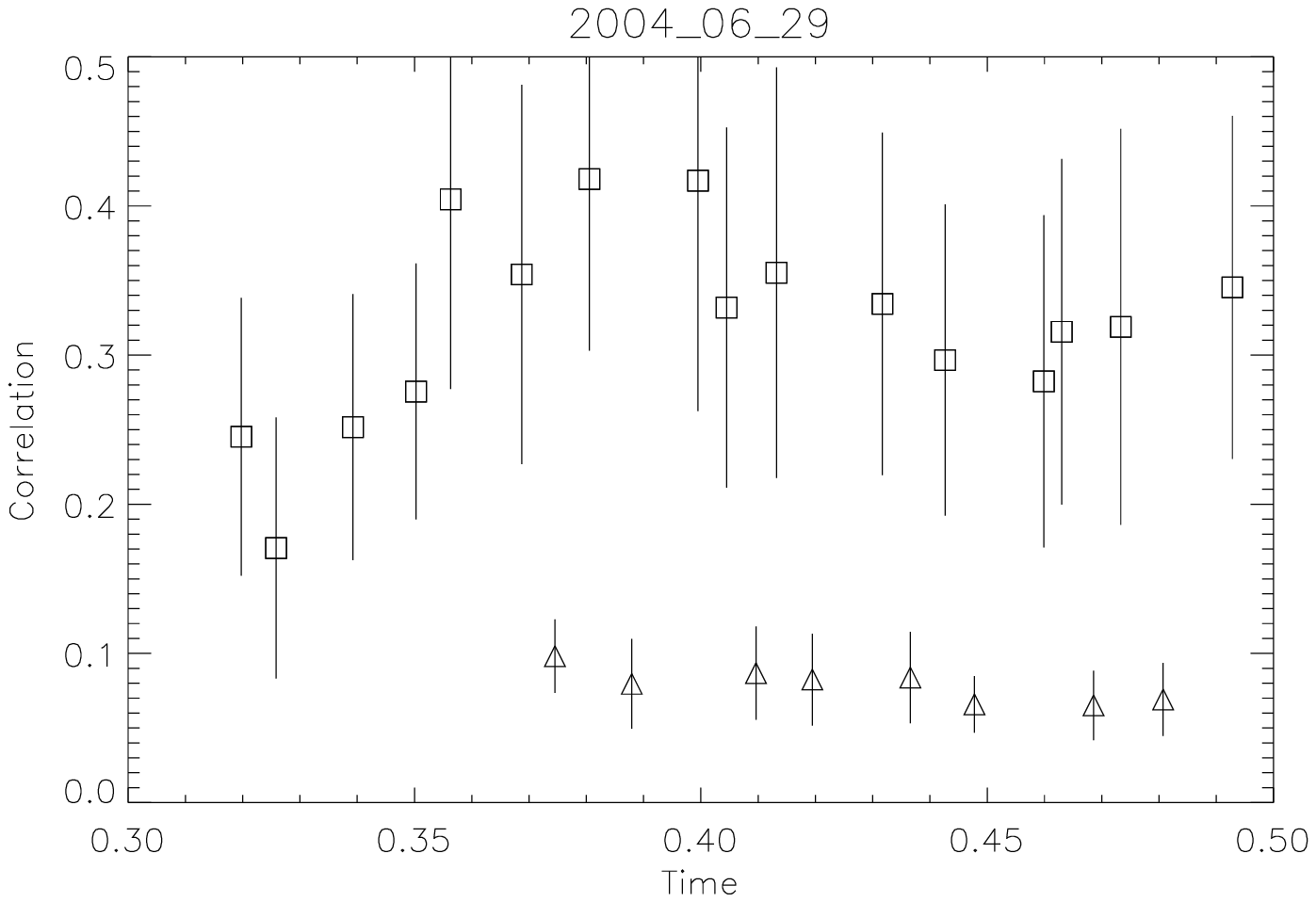}}
\caption{Two examples of a complete measurement sequence using the calibrator
HD 197373 (squares) and the object HD 203280 (triangles) 
on a night of good seeing
(top) and a night of poor seeing (bottom). These data are from the same two nights as those 
shown in Figure \ref{fig_vhist}. \label{fig_cal}}
\end{figure}

Figure \ref{fig_cal} shows two complete measurement sets taken from the
same two nights as shown in Figure \ref{fig_vhist}. The projection angle
on the object was different on these two nights so one expects a change
in the object visibility, indeed the actual object visibility was higher
during the night of poor seeing than during the night of good seeing.
Still, the calibrator raw correlation changes significantly in poor
seeing and remains fairly stable in good seeing. The bottom line in
interferometry is calibration of the data and the biggest impact on this
is the seeing. Despite our efforts to de-bias our correlation estimators,
the formal errors found in the measurement process are always smaller
than those created by the calibration process. Our best efforts during
the observing season of 2004 resulted in calibrations good to 4\% when
using the open air beam combiner without spatial filtering. This is 
reduced to better than 1\% when using the fiber based combiner and 
spatial filtering.

As of this writing, more than 5000 data sets have been collected from 12 
baseline pairs since late 1999. The majority of the early observations have 
been for engineering purposes, and it has only been since the spring of 
2002 that the observing has been directed toward science rather 
than engineering. The first full year of observing directed primarily
toward science observations was 2004. In that year we opened the
telescope domes on a total of 229 nights, and of those nights we
collected useful data on 154 nights. While seeing clearly affects our
ability to calibrate the data we have found it is still possible to
collect data until the correlation length $r_0$ falls below 3cm. Below
this it is still possible to obtain fringes, though it is all but
impossible to achieve any sort of worthwhile calibration. No strong
correlations between data quality and time of night have been found,
though we have found four distinct observing seasons here on Mount
Wilson. During the winter, the seeing is poor and we are frequently
closed due to high winds, rain or snow. Conditions improve
during the spring and our best season occurs during the summer months
where we often have many clear nights of excellent seeing in a row. The
fall season is similar to the spring, though often affected by high
winds.

\section{Conclusion}

The full-time science program of the CHARA 
Array has been underway for only a year, and we expect many exciting
years of science, technical development and data analysis exploration
to come. 
The CHARA Array was built within the original budget and on 
a reasonable time line by a small university-based group of scientists. 
Many lessons have been learned along the way from the earliest days of CHARA 
that might be valuable to future projects, and so we plan to describe that
experience in another venue.

All six telescopes, 
light pipes, delay lines, beam reducers, longitudinal dispersion 
correctors and beam sampling systems are installed and fully operational, 
and the Array is now regularly scheduled for a variety of astronomical 
projects. While routine science observing is underway, we look forward to 
continual technical and performance improvements including such capabilities 
as phase closure and multi-way beam combination expected to begin in 2005.

\acknowledgments

Construction funding for the CHARA Array has been provided by the 
National Science Foundation through grant AST~94-14449, 
the W. M. Keck Foundation, the David and 
Lucile Packard Foundation, and by Georgia State University. We are also
grateful to the National Science Foundation for its nurturing of the 
project through separate Phase A and Phase B grants. 
The construction of our exhibit hall was made possible
by a gift from Mr. Jack R. Kelly to Georgia State University.
The AROC is named after Georgia State's former Vice President for
Research, Cleon C. Arrington, in 
recognition of his support of CHARA over the years. 
We wish to thank Robert Jastrow, former director of 
the Mount Wilson Institute for his hospitality 
and cooperation. We similarly acknowledge the cooperation of Terry Ellis, 
former District Ranger for the Los Angeles River Ranger District of 
the Angeles National 
Forest, in working with CHARA to develop this project on National Forest 
lands. We also thank E.J. Simison of Sea West Enterprises for his enthusiasm
and initiative in the design and construction of the Array facilities, as well 
as Gale Grant for his work on the dome automation.

\clearpage

\begin{deluxetable}{crrrr}
\tablewidth{0pt}
\tablecaption{Available baselines. These numbers are based on the
results of a global baseline solution from mid-2004. The RMS error
of this fitting process was 1891-$\mu$m.\label{table_baselines}}
\tablehead{
\colhead{Telescopes} &
\colhead{East (m)} &
\colhead{North (m)} &
\colhead{Height (m)} &
\colhead{Baseline (m)}
}
\startdata
S2-S1 & -5.748     & 33.581     & 0.644    &   34.076       \\
E2-E1 & -54.970    & -36.246    & 3.077    &   65.917       \\
W2-W1 & 105.990    & -16.979    & 11.272   &   107.932      \\
W2-E2 & -139.481   & -70.372    & 3.241    &   156.262      \\
W2-S2 & -63.331    & 165.764    & -0.190   &   177.450      \\
W2-S1 & -69.080    & 199.345    & 0.454    &   210.976      \\
W2-E1 & -194.451   & -106.618   & 6.318    &   221.853      \\
E2-S2 & 76.149     & 236.135    & -3.432   &   248.134      \\
W1-S2 & -169.322   & 182.743    & -11.462  &   249.392      \\
W1-E2 & -245.471   & -53.393    & -8.031   &   251.340      \\
W1-S1 & -175.071   & 216.324    & -10.818  &   278.501      \\
E2-S1 & 70.401     & 269.717    & -2.788   &   278.767      \\
E1-S2 & 131.120    & 272.382    & -6.508   &   302.368      \\
W1-E1 & -300.442   & -89.639    & -4.954   &   313.568      \\
E1-S1 & 125.371    & 305.963    & -5.865   &   330.705      \\
\enddata
\end{deluxetable}

\clearpage

\clearpage

\clearpage

\clearpage

\clearpage

\clearpage

\clearpage

\clearpage

\clearpage

\clearpage

\clearpage

\clearpage
\end{document}